\documentclass[conference]{IEEEtran}
   \usepackage{amsmath,amsfonts,amssymb}
   \usepackage{epsfig,url}
   \usepackage{graphicx,color,footmisc}
   \usepackage{subcaption}
   \usepackage{algorithm}
   \usepackage{tikz}
   \usepackage{color}
   \usepackage{cite}
   \usepackage{epstopdf}

   \title{A Framework for Wireless Broadband Network for Connecting the Unconnected}
   \author{
   	\IEEEauthorblockN{Meghna Khaturia, Prasanna Chaporkar and Abhay Karandikar}
   	\IEEEauthorblockA{Department of Electrical Engineering, Indian Institute of Technology Bombay, Mumbai-400076 \\ Email: \{meghnak,chaporkar,karandi\}@ee.iitb.ac.in}}
   
\begin{document}
   	\maketitle
	
	\begin{abstract}	
		A significant barrier in providing affordable rural broadband is to connect the rural and remote places to the optical Point of Presence (PoP) over a distances of few kilometers. A lot of work has been done in the area of long distance Wi-Fi networks. However, these networks require tall towers and high gain (directional) antennas. Also, they work in unlicensed band which has Effective Isotropically Radiated Power (EIRP) limit (e.g. $1$ W in India) which restricts the network design. In this work, we propose a Long Term Evolution-Advanced (LTE-A) network operating in TV UHF to connect the remote areas to the optical PoP. In India, around $100$~MHz of TV UHF band IV ($470$-$585$ MHz) is unused at any location and can be put to an effective use in these areas~\cite{Gaurang}. We explore the idea of multi-hop topology for the proposed network. We also compare the performance of multi-hop network with the Point to Multipoint (PMP) topology. The results show that multi-hop network performs much better than the PMP network. We then formulate a Linear Programming (LP) problem of generating optimal topology and compare its performance with the multi-hop network. Overall, the analysis implies that an optimally planned LTE-A network in TV UHF band can be a potential solution for affordable rural broadband.
	\end{abstract}
	
		\section{Introduction}
	In spite of the spectacular growth in Internet usage, a staggering 47\% of the global population is still unconnected~\cite{unbb}. In the Indian context, the number of broadband subscribers is only $291$ million (as on $31^{st}$ May, $2017$) out of a total population of $1.34$ billion~\cite{TRAI}. The state of broadband is even worse in the rural areas of India. Providing broadband connectivity in these areas is a very challenging task due to reasons that include i) lack of fiber infrastructure, ii) low Average Revenue Per User (APRU), and iii) difficult terrain.  The Government of India under its BharatNet \cite{bharatnet} initiative aims at providing Points of Presence (PoPs) with optical connectivity to local self-government offices at village level called Gram Panchayats (GPs) in India. There are $250,000$ GPs in India serving about $640,000$ villages. Currently, fiber cable has been laid for about $100,000$ GPs~\cite{bharatnet}. The remaining GPs and their neighboring villages are still unconnected. Taking optical fiber to each and every village or GP is very time consuming and expensive. Other options like Digital Subscriber Line (xDSL), coaxial cable are infeasible because these are difficult to deploy and have distance limitations. Thus, there is a need to look for  wireless alternatives so as to efficiently connect these villages. 
	
	In the last decade, there have been several initiatives around the world to provide broadband to the rural communities. AirJaldi~\cite{airjaldi} is a private enterprise that is working towards providing broadband in rural areas of India. Ashwini~\cite{bRamanrural} and Aravind~\cite{patra} are some other examples of rural network testbeds deployed in India. Hop-Scotch is a long distance Wi-Fi network based in UK~\cite{darbari}. LinkNet is a 52 node wireless network deployed in Zambia~\cite{Zambia}. Also, there has been a substantial research interest in designing the long distance Wi-Fi based mesh networks for rural areas~\cite{bRaman, rethinking, ruralmesh}. All these efforts are based on IEEE 802.11 standard \cite{802.11} in unlicensed band i.e. $2.4$ GHz or $5.8$ GHz. When working with these frequency bands over a long distance point to point link, Line of Sight (LoS) may be required between transmitter and receiver. To establish LoS, typically, towers of sufficient height and high gain directional antennas could be necessary infrastructure requirement. Maintenance of such towers is also very difficult in remote areas. This increases the overall cost of the network significantly even while Wi-Fi devices have been highly commoditized. The unlicensed band has other problems such as  Effective Isotropically Radiated Power (EIRP) limit (e.g. $1$ W in India). Covering longer distances with high throughput may become a difficult task with such EIRP limits. 
	
	In this paper, we propose a wireless \textit{middle mile} network for connecting an optical PoP to all the villages where the Internet connectivity is required. The end users in these  villages can access the Internet through Wi-Fi Access Points (APs) which are backhauled over the middle mile network. We propose the deployment of  Long Term Evolution - Advanced (LTE-A) in TV UHF band as the middle mile network. Owing to its good long distance propagation characteristics, the links of middle mile network can work in non LoS or near LoS fashion and thus the UHF band can be put to a great use in rural areas. 
	
	We evaluate  multi-hop topology as well as Point to Multi-point (PMP) network topology for the middle mile. We first begin by designing the multi-hop topology without considering the load requirement of the nodes in the network. We then formulate the problem  of generating an optimal topology which takes into consideration the load requirement of the nodes within the framework of Linear Programming (LP). The results of the LP based solution are compared with the PMP and multihop topologies. The solution of LP problem is able to satisfy the load requirements of the nodes in significantly many cases as compared to multi-hop. These results show that if we optimally plan the network for rural areas, we can meet the requirements in much more efficient way.
		
	The rest of the paper is organized as follows. Section II describes the framework of the middle mile network. In Section III, we compare the performance of various network topologies. Section IV deals with the problem formulation of generating an optimal topology which satisfies the network load. Section V concludes our work. 
	
	


	\section{Middle Mile Network Architecture}
	
\begin{figure}[!ht]
		\centering
		\includegraphics[scale = 0.6]{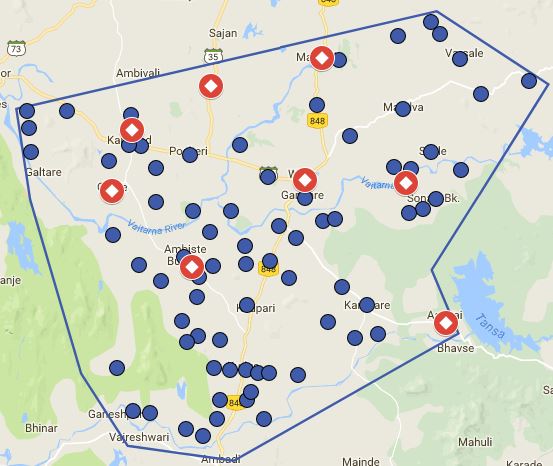}
		\caption{Location of GPs (Round) and Location of towers (Diamond) for Wada Block in Palghar, Maharashtra, India. This map is available online at the link provided in~\cite{map}.}
		\label{fig:wada}
	\end{figure}

	We consider an example of a typical rural setting in India to gain an insight into the network infrastructure present in rural unconnected areas. For example, a rural area named Wada located in the state of Maharashtra in India, is spread over an area of $800$~sq. km and comprises of $85$ GPs. As depicted in Fig.~\ref{fig:wada}, there are only $8$ towers (approx. 40 m high) in this area. These towers provide only voice connectivity to the GPs located within a radius of around $2-3$~km. Assuming that the optical PoP is present at these towers, we have to connect all the GPs in this area. The minimum distance between a GP and the nearest tower is around $500$ m, while the maximum distance is around $10-12$~km.  With no tall tower available at the GP or village location, it is not possible to set up a point to point radio link to serve these villages. This scenario has motivated us to explore solutions such as the use of TV UHF band for connecting these unconnected GPs and their villages with the available infrastructure only.
	
	Even the connectivity requirements of rural areas are very different from urban areas and this calls for a different approach while designing a network for rural areas. Unlike urban areas, the purchasing power is low in rural areas and hence it is important to look for a low cost solution for these areas. The technology for rural areas need not cater to high mobility when primary fixed connectivity is missing. What we really need for rural areas is a i) robust, ii) energy efficient and iii) easy to manage rural broadband technology. We now discuss the architecture of the middle mile network which has been designed taking these requirements and the available infrastructure into consideration.
	
\begin{figure}
			\centering
			\includegraphics[scale = 0.27]{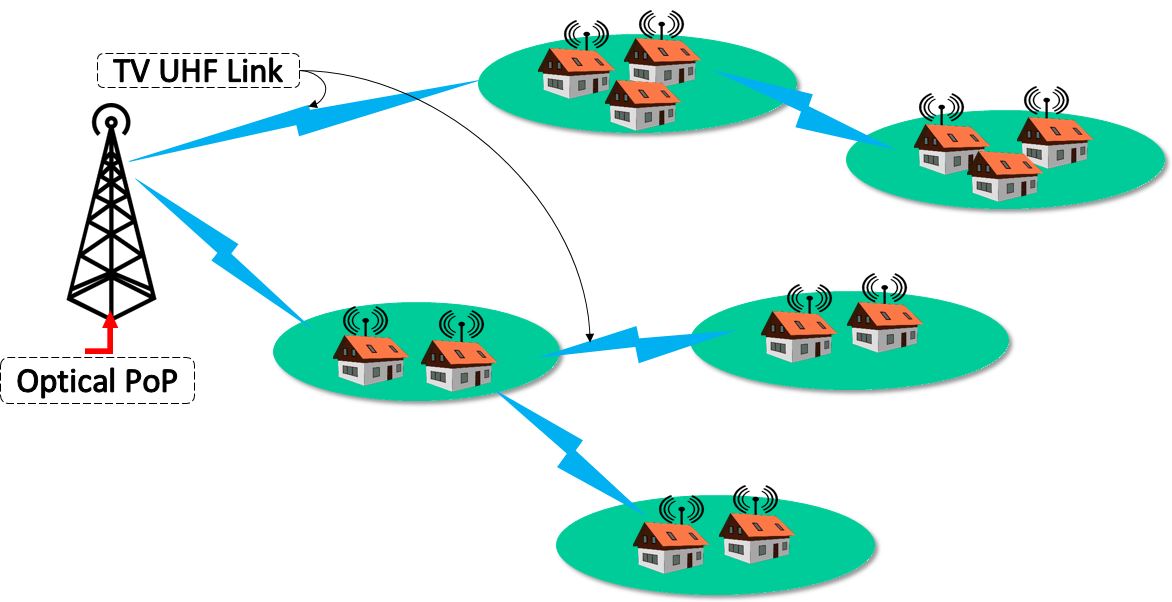}
			\caption{Middle mile network}
			\label{fig:middlemile}
		\end{figure}

	A middle mile network is illustrated in Fig.~\ref{fig:middlemile}. A low cost broadband access to the end-users in the villages can be provided via Wi-Fi APs. To backhaul these Wi-Fi APs, we explore an LTE-A based middle mile network in TV UHF band. An LTE-A eNB is located at the tower and locally connects to the optical PoP present at the tower. The LTE-A UEs connect to the eNB via TV UHF band and locally connect to  the Wi-Fi APs in the villages. 
				
	Moreover, the TV UHF band is highly underutilized in India and has very good propagation characteristics. A detailed quantitative assessment of TV White Space in India reveals that 12 out of 15 channels of TV UHF Band IV ($470$-$585$ MHz) are available at any location~\cite{Gaurang}. 
	In our previous work~\cite{IEEEcommmag}, we set up a testbed operating in TV UHF band which demonstrated how effectively this band can aid in reaching the far flung areas. While performing experiments on our testbed, we were able to cover around 5 km in a Non Line of Sight manner which is not possible at higher frequencies. Thus, employing this band in the middle mile network will significantly reduce the infrastructure cost. The need to erect tall towers is immensely reduced owing to the large outreach of this band even at low radio powers (1 W). Consequently, the power consumption can be lowered which makes a case for using solar energy. The deployment of devices working in TV UHF band can be done with ease as line of sight connectivity is not a crucial requirement. This feature is extremely helpful in areas with hilly terrain. Also, the foliage loss in this band is lower as compared to higher frequencies.

		
	Selecting an optimal network topology is a very important step in designing the middle mile network. The Point to Multipoint (PMP) topology is a most popular approach of connecting an eNB to its UEs. This network topology might not be suitable for rural areas since they are very sparsely populated and need connectivity only at certain locations. The inhabitants in these areas live in a clustered manner. These clusters can be a few kms away from each other. Thus there is a need to explore other topologies such as multi-hop to reach the far flung places. In this paper, we have considered multi-hop topology with limited number of hops. We form this multi-hop topology with the help of available infrastructure and short towers (10m) or the rooftops of a two storey building. There are several benefits of using multihop topology which can be listed as:
	
		\begin{itemize}	
			\item This topology will help in reaching those clusters/users which are not accessible by single hop.
			\item It will also be helpful in overcoming obstructions such as hills by forming a wireless hop over it.
			\item Much larger capacity can be supported with this topology due to spatial reuse of the resources.
			\item As we are limiting the multihop links on rooftops or short towers, this will help in significantly reducing the cost of the network by eradicating the need of tall towers. 
		\end{itemize}
		
	\section{Performance Comparison of Multi-hop and Point to Multipoint (PMP) Topology}
In this section, we will analyse the performance of PMP and multi-hop topology under the system model described below:
	\subsection{System Model}
	 \label{system}
	 Consider an optical PoP located randomly in an area of $L$x$L$ sq. km. There are $N$ number of Wi-Fi APs in this area which needs to be backhauled. These Wi-Fi APs provide broadband access to the end users in the rural areas. Hence, these Wi-Fi APs are set-up depending upon the location of the habitat clusters in the rural areas and will be distributed randomly over the considered area.  We have assumed random load at the Wi-Fi APs. 
	 To backhaul these Wi-Fi APs, we consider an LTE-A based network operating in TV UHF band. The eNodeB (eNB) is located at the optical PoP while the User Equipments (UEs)/Relay Nodes (RNs) are co-located with the Wi-Fi APs. As we are considering multi-hop network, a node may have to relay the traffic for the other nodes. We facilitate this with RNs as defined in the LTE-A standard. Here, we consider a fixed wireless network i.e. UEs are stationary. 

	 We assume that a part of TV UHF band is available which will be used in our wireless broadband network. The available band is divided into Resource Blocks (RBs) as specified in LTE-A standard. Each RB here is $180$~KHz long in frequency and $0.5$~ms in time. The resource allocation to the UEs/RNs is done based on the demand generated at each Wi-Fi AP. 
	 We have considered only downlink traffic i.e. from PoP to the Wi-Fi APs. Next, we will discuss the system characteristics that are specific to the topology.

		\subsubsection{Point to Multipoint (PMP) Topology}
\begin{figure}[!ht]
		  \centering
		  \begin{subfigure}[b]{0.4\linewidth}		 
			  \centering
			  \includegraphics[trim={1cm 1cm 1cm 1cm},scale = 0.28]{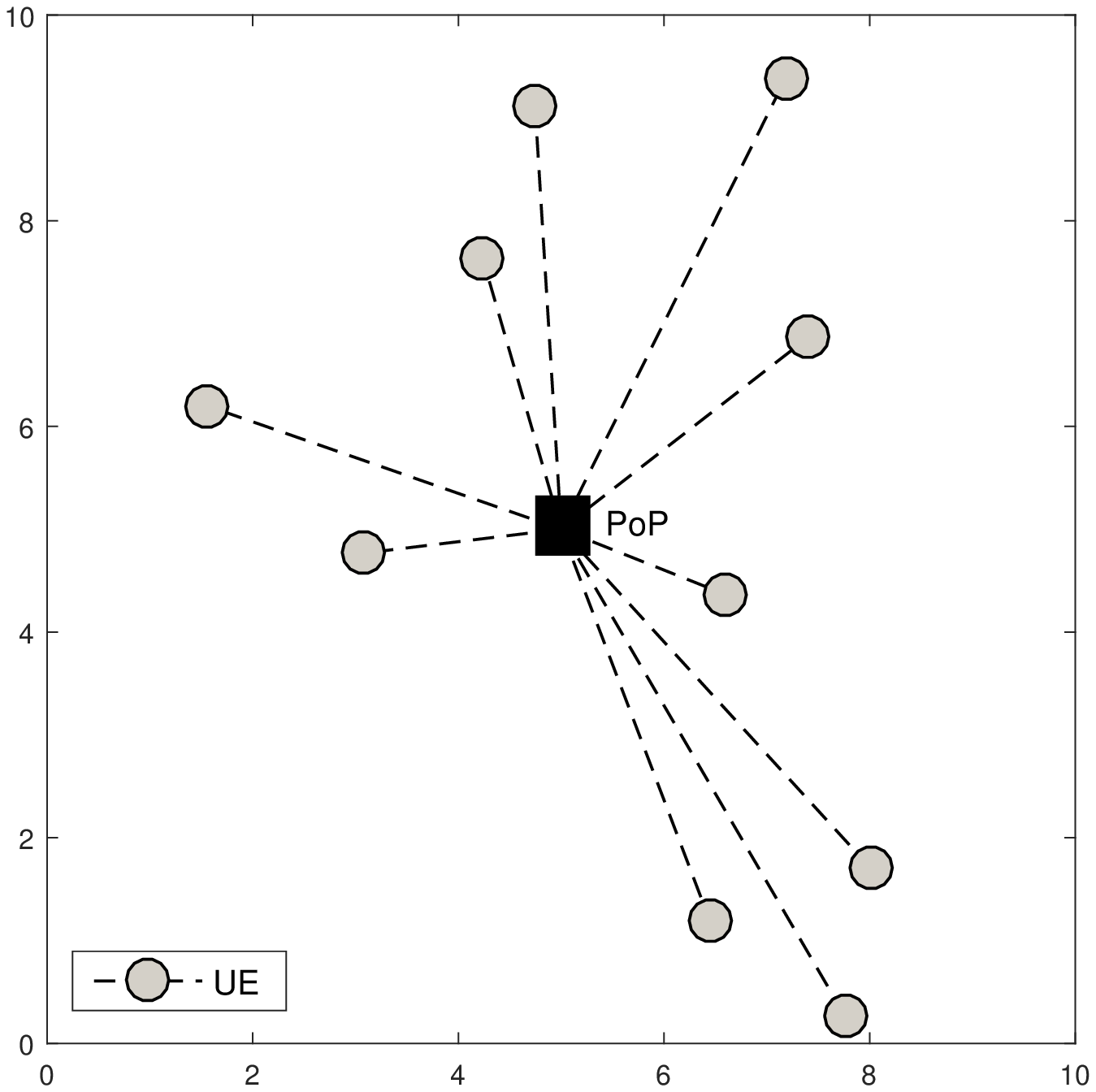}
			  \caption{PMP topology }
			  \label{fig:PMP}		  
		  \end{subfigure}
		  \quad \quad
		  \begin{subfigure}[b]{0.4\linewidth}
			  \centering
			  \includegraphics[trim={1cm 1cm 1cm 1cm},scale=0.28]{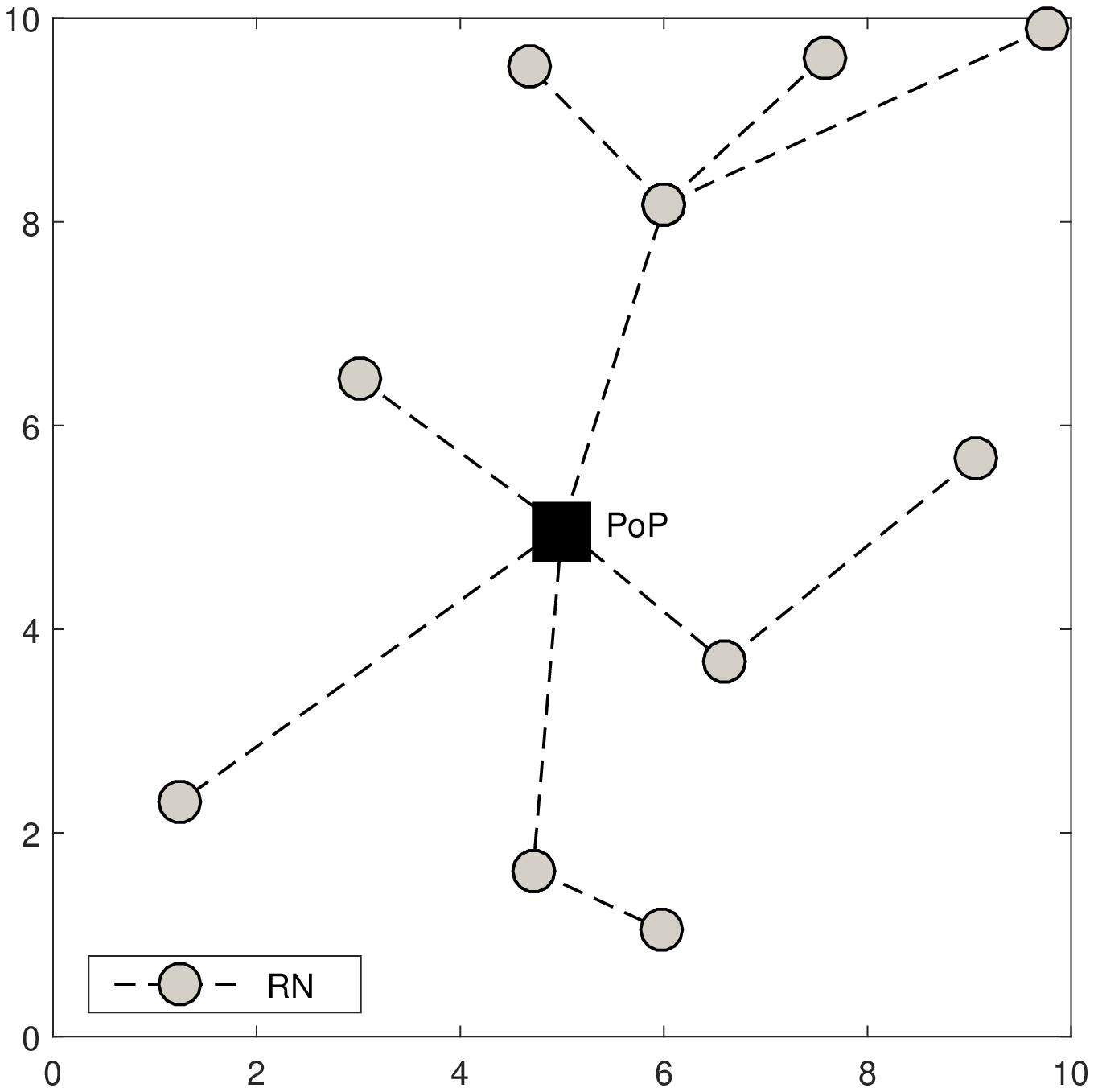}
			  \caption{Multihop topology }
			  \label{fig:multihop}
		  \end{subfigure}
		  \caption{Example of Network Topology (axis in km)}
		\end{figure}

		The eNB radiates through an omni-directional antenna which is mounted on a tall tower (approx $40$~m). A UE is preferably mounted at the rooftop of a house. An example topology is shown in the Fig.~\ref{fig:PMP}. In a PMP topology, multiple UEs connects to the eNB. The resource allocation is done directly depending on the demand generated at Wi-Fi APs.

	%

		\subsubsection{Multihop Topology}
		Similar to the PMP topology, eNB is also mounted on a tall tower in multi-hop topology. This eNB connects to the RNs in a multi-hop manner as shown in the Fig.~\ref{fig:multihop}. Note that we have only considered tree topology and not mesh. These RNs not only serve the co-located Wi-Fi AP but also relay the traffic for other Wi-Fi APs. Here, all the RNs in the network have directional antennas. If there exist a link between two RNs, then there is a dedicated transmit - receive radio pair for that link. An RN works in full duplex mode i.e. it can receive as well as transmit simultaneously. Full duplex radios are very costly to enable the full duplex mode, so we consider that an RN has separate antennas for transmission and reception. Also, a relay node can connect to limited number of relays as a short tower/pole may not be able to support more than a few equipment. We also limit the maximum number of possible hops as it will increase the delay in the network.
		
		Next, we explain how the topology formation and resource allocation is carried out in multi-hop network. 
	
		  \begin{itemize}
		  \item We have used Minimum Weight Spanning Tree (MWST) algorithm to design the multihop network. Given a connected, weighted graph, the MWST algorithm will find an acyclic graph with minimum possible weight connecting all the vertices together. We have used Prim's MWST algorithm and modified it to limit the maximum number of hops and maximum allowed degree. The maximum possible hops can be either $2$ or $4$. The degree of a node in a multi-hop network can atmost be $4$. The weight in the graph is the distance metric. This metric was selected to avoid any long distance link which will bring down the throughput performance of the network. 
		  \item To allocate RBs in the multi-hop network,we need to calculate the required capacity of each link. The capacity required at any link is the sum of the required load of all Wi-Fi APs served under that branch. This obtained link capacity is converted into the number of RBs required at each link. The conversion from load to required RBs is a function of the received Signal to Noise Ratio (SNR) at each link. We use a 3GPP Rural Macrocell (RMa) model to calculate path loss for every link in the network~\cite{TS36.814}. As these links are directional, the interference to other links will be very less. However, those links which originate from the same node will interfere heavily due to back lobes. So, the links originating from same node will be allocated mutually exclusive RBs. The resource allocation problem under the above scenario is similar to the edge coloring problem with multiple colors. Here, colors are the number of RBs available in the network. After solving the edge coloring problem on this graph, we get the resource allocation.
		  \end{itemize}
	
	Our objective is to design the optimal topology for the network. Before formulating the optimal topology problem, it is important to look into the performance of both the topologies. This performance comparison is important in understanding the need of exploring new topologies for rural broadband network.
	
	\subsection{Simulations Results}
			\begin{table}
				\centering
				\caption{Multi-hop Network Simulation Parameters}
				\label{tab:multihop}
				\renewcommand{\arraystretch}{1.2}
				\begin{tabular}{|l|l|}
					\hline
					\textbf{Parameters}                             & \textbf{Values}   \\ \hline \hline
					Frequency Band                        			& $500$-$520$ MHz         \\ \hline   
					Transmit Power 			                        & $27$ dBm             \\ \hline
					eNB Antenna  		               				& Directional (Multi-hop)  \\ 
					& Omni-directional (PMP) \\ \hline
					UE/RN Antenna            		         		& Directional (Multi-hop)             \\
					& Omni-directional (PMP) \\ \hline
					eNB Antenna Gain  			              		& $10$ dBi               \\ \hline
					UE/RN Antenna Gain  		                	& $10$ dBi              \\ \hline
					eNB Height  			          				& $30$ m              \\ \hline
					UE/RN Height	          		      			& $10$ m               \\ \hline
					Deployment Area                                 & $100~km^2$ \\ \hline
					Channel Model									& NLoS Path Loss for \\ & Rural Areas \cite{TS36.814}\\ \hline  
					Number of Deployment 			                & $2000$             \\ 
					Scenarios & \\ \hline
				\end{tabular}	
				\vspace{0.25cm}
			\end{table}
	
			\begin{figure}
				\centering
				\includegraphics[trim={1cm 0.5cm 1cm 1cm}, scale = 0.5]{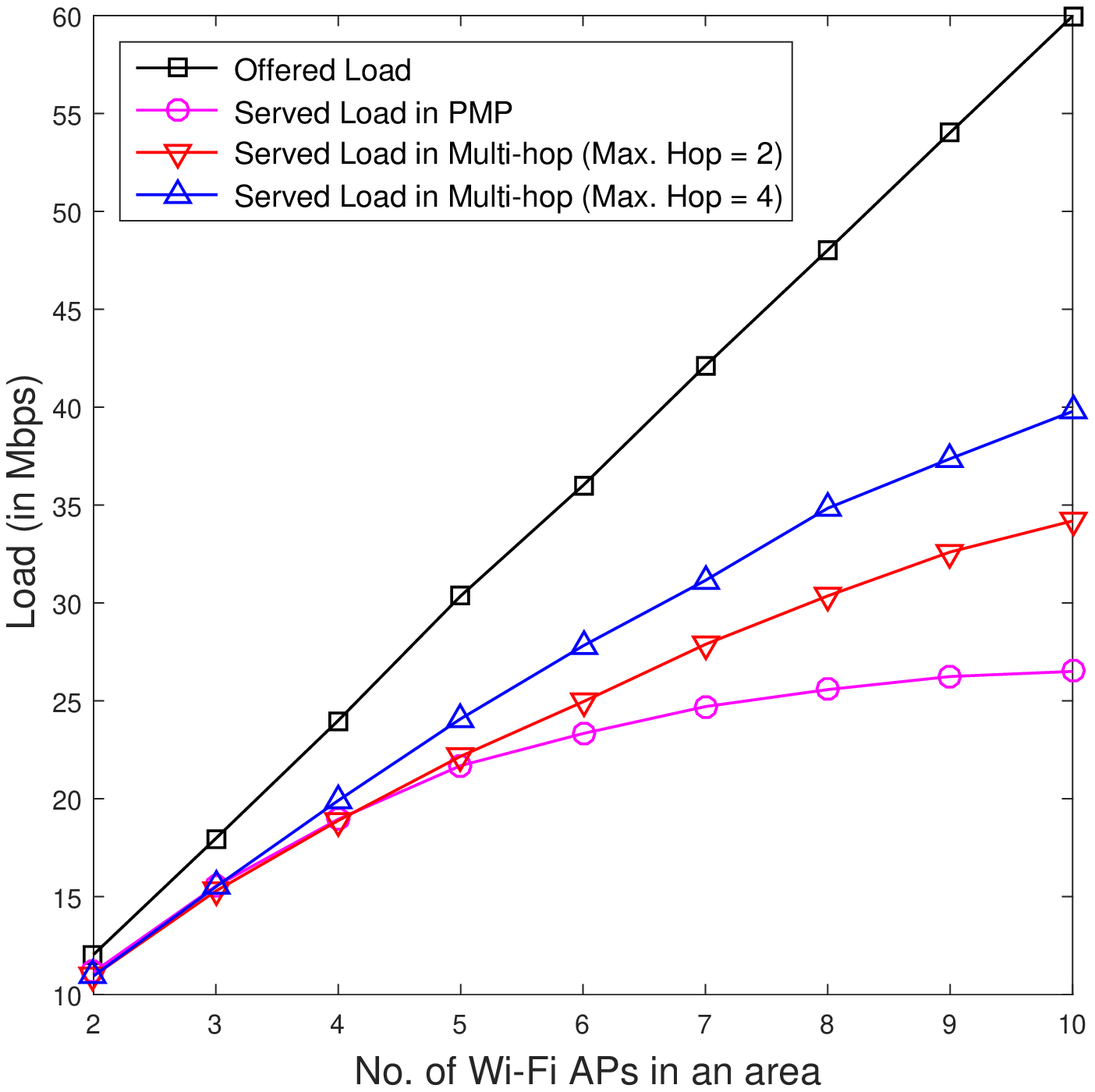}
				\caption{Served load under various topologies vs the number of Wi-Fi APs in an area}
				\label{fig:load}
			\end{figure}			
			\begin{figure}
				\centering		
				\includegraphics[trim={1cm 0.5cm 1cm 1cm}, scale = 0.5]{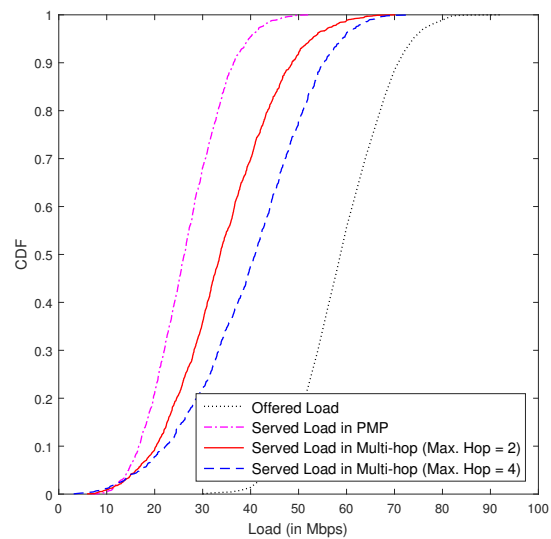}
				\caption{CDF of the served load in the network of 10 Wi-Fi APs in an area}
				\label{fig:1}
			\end{figure}	
			\begin{figure}
				\centering	
				\includegraphics[trim={1cm 0.5cm 1cm 1cm}, scale = 0.5]{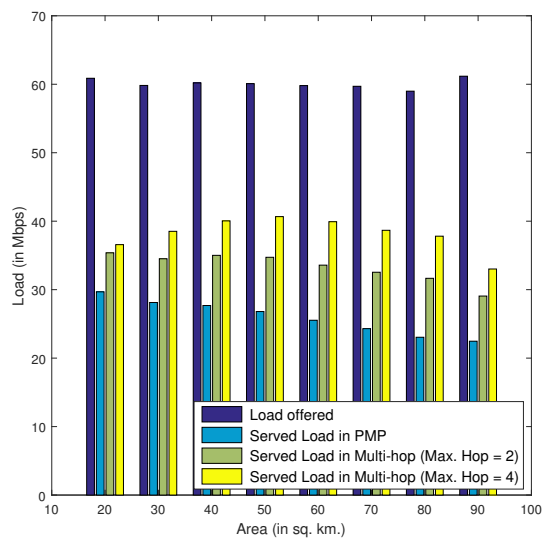}
				\caption{Variation of served load with reference to increase in area of the network of 10 Wi-Fi APs}
				\label{fig:2}
			\end{figure}
		We have simulated the middle mile network in an area of $10$~km X $10$~km. The location of optical PoP and Wi-Fi APs are randomly generated. The load at Wi-Fi APs is randomly picked from a set of $\{2,4,6,8,10\}$~Mbps. We have only considered downlink traffic in these simulations. The PMP and the multihop network are formed in the same way as described in the previous section.  We assume that a band of 20 MHz is available in TV UHF band. This band is equivalent to 100 RBs in an LTE network. We will allocate these RBs in the network based on the demand generated at Wi-Fi APs. 
		The simulation parameters for PMP and multihop are mentioned in Table~\ref{tab:multihop}. The simulations are done in MATLAB. 
	
		Now, we vary the number of Wi-Fi APs in the network from $2$ to $10$. The comparison between the PMP and multi-hop network is done with reference to the served load in the middle mile network. The served load is the sum of downlink throughput served to all the Wi-Fi APs in the network. As shown in the Fig.~\ref{fig:load}, the multi-hop network performs better in terms of the served load. As we allow more number of hops to $4$, the served load of the network increases. This shows that multi-hop networks are very beneficial in this scenario.  
	
		Next, consider a network of 10 Wi-Fi APs in an area of $10$~km X $10$~km. We will now analyze the probability distribution of the served load under PMP and multi-hop topologies. As the Fig.~\ref{fig:1} shows, multi-hop topology performs much better than the PMP topology. We also observe an increase in performance with the increase in maximum allowed hops from $2$ to $4$. 
		
		As shown in the Fig.~\ref{fig:2}, we have also observed the performance of these topologies with reference to increasing in area. It is interesting to note that the performance of multi-hop remains almost same throughout while the performance of PMP decreases with the increase in area.  
		
		Overall, the multi-hop topology prove to be a potential solution in rural areas. However, we can do much better if we take the required load into consideration while designing the network. We will explore this problem next.
		
	
\section{Optimal Topology Design}
	In the previous section, we discussed the performance of various topologies. However, we did not consider the load at the Wi-Fi APs while designing the network. This is a suboptimal way to design the network. In this section we will formulate the problem of optimal network design and resource allocation.  

		\subsection{Problem Formulation}
		As described in Section~\ref{system}, an eNB is co-located with the optical PoP and there are $N$ Wi-Fi APs/RNs randomly distributed in an area of $L$ x $L$ sq. km. Consider a complete network graph consisting of an eNB and $N$ RNs. Hence, there are $N+1$ nodes in the network graph. We will denote a node in the network by $K_i$ where
			\begin{equation*}
				K_i = \left\{\begin{matrix}
				\text{eNB} & i=0, \\  
				\text{RN}_i & otherwise.
				\end{matrix}\right.
			\end{equation*}
		
		Every link between node K$_i$ and node K$_j$ has a capacity $C_{ij}$. Every node K$_i$ has a load requirement denoted by $\lambda_{i}$ except K$_0$ i.e  $\lambda_{0} = 0$. The objective is to find a sub-graph which satisfies the load requirements of the Wi-Fi APs in the network. Here, we want to minimize the link utility $\beta_{ij}$ over all the links  which will aid in selection of those links which have higher capacity $C_{ij}$. Consequently, the number of links in the obtained sub-graph will be minimum. Mathematically the optimization problem can be stated as follows:
		
			\begin{equation}
				\label{eq:obj}
				 \underset{\beta_{ij}} \min \quad \sum\limits_{i=0}^{N} \sum\limits_{j=0}^{N}\beta_{ij}, 
			\end{equation}
			\begin{equation}
				\text{subject to}   \quad \lambda_{i} + \sum\limits_{j=0}^{N}\beta_{ji}C_{ij} \leqslant \sum\limits_{j=0}^{N}\beta_{ij}C_{ij} \quad \quad \forall \ i \neq 0,	
				\label{eq:flow}	
			\end{equation}			
			\begin{equation}
				\sum\limits_{j=0}^{N}\beta_{ij} \leqslant 1 \quad \quad \forall \ i,
				\label{eq:util1}	
			\end{equation}
			\begin{equation}
				\sum\limits_{i=0}^{N}\beta_{ij} \leqslant 1 \quad \quad \forall \ j.
				\label{eq:util2}	
			\end{equation}
		
		Constraint Eq. (\ref{eq:flow}) conserves the flow at every node in the network except at node K$_0$ i.e. eNB. As the eNB is the source of the flow in the network, flow conservation can not be applied. A node K$_i$ should not utilize more than the available resource which is taken care by Eq.  (\ref{eq:util1}) and (\ref{eq:util2}).  
		
		The above mentioned optimization problem is a Linear Programming (LP) problem which can be solved by an LP solver. After obtaining the link utility, we multiply the obtained utility value by available RBs. This solution not only gives us the network design but also the resource allocation on every link. We will now discuss the simulation results.

		\subsection{Simulation Results}
		We have simulated the network in an area of $10$~km X $10$~km. The location of optical PoP and Wi-Fi APs are randomly generated. The load at Wi-Fi APs is randomly picked from a set of $\{2,4,6,8,10\}$~Mbps. We have only considered downlink traffic in these simulations. We assume that a band of 20 MHz is available in TV UHF band. The LP optimization problem may or may not be able to give a feasible topology for every deployment scenario. Therefore, we have only considered feasible topologies i.e. those topologies for which the load requirement of all the Wi-Fi APs is satisfied. We have compared the results of the results of the LP problem with the topologies considered in Section III. We do maintain the consistency and take only those network topologies which are feasible in PMP and multi-hop network while comparing the performances. 
		
\begin{figure}
			\centering
			\includegraphics[trim={1cm 0.5cm 1cm 1cm}, scale = 0.5]{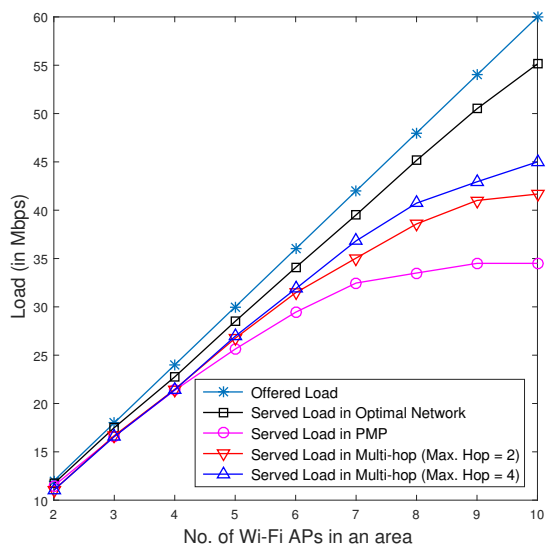}
			\caption{Served load under various topologies vs the number of Wi-Fi APs in an area}
			\label{fig:load_opti}
		\end{figure}

		We have simulated the network for the same parameters as given in Table~\ref{tab:multihop}. Fig.~\ref{fig:load_opti} shows that we are able to serve much higher load as compared to the PMP and multi-hop topologies. The topology obtained by solving the LP problem can be a mesh topology but not a very dense mesh network. 
		
\begin{figure}
			\centering		
			\includegraphics[trim={1cm 0.5cm 1cm 1cm}, scale = 0.5]{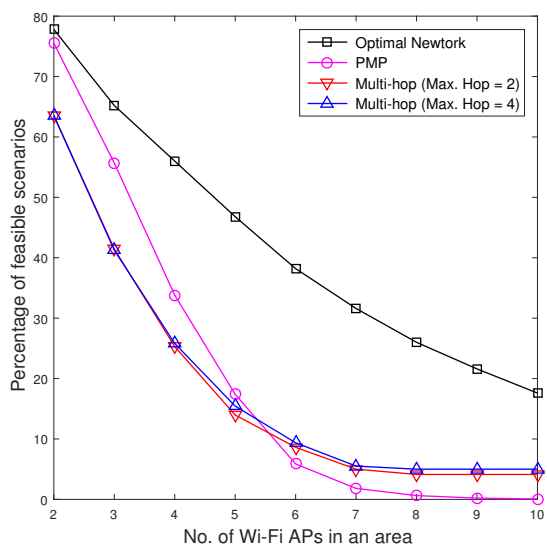}
			\caption{Percentage of feasible deployment scenarios under various topologies vs the number of Wi-Fi APs in an area}
			\label{fig:feasible}
		\end{figure}

		It is also important to observe that the number of feasible deployement scenarios under each topology. Fig.~\ref{fig:feasible} shows that the performance of the LP problem is much better than the other topologies. It can also be observed that the percentage of feasible topologies in PMP is less than multi-hop as we increase the density of graph. 
		
			
	\section{Conclusion}
	     In this paper, we address the issue of providing broadband connectivity to remote underserved areas. Specifically, we investigate an architecture for fixed broadband. In this architecture, an LTE-A network operating in TV UHF band serves as  the wireless middle mile for backhauling Wi-Fi hotspots. Towards this, we explore multi-hop topology  and compare its performance with the PMP network. To further optimize the network design, we take into consideration the load requriements of the network and propose an optimal design within the framework of linear programming. Our results demonstrate that the optimal network performs much better than the multi-hop and PMP network. We conclude that exploring new efficient network topologies can be an important step in desiging the middle mile network. We intend to experimentally verify these outcomes on our live testbed \cite{IEEEcommmag}.

	\nocite{}	
	\bibliographystyle{ieeetr}
	\bibliography{myrefs}

\end{document}